**Nonlinear trajectories of lung function recovery in patients with pulmonary disease: empirical evaluation of longitudinal modeling approaches**

Zongyue Teng[1], Ningkun Zhou[1], Xinyu Zhang[1], Robert Wallis[2,3], Qingyan Xiang[1,3,*]

[1] *Department of Biostatistics, Vanderbilt University Medical Center, Nashville, TN, USA*

[2] *Department of Medicine, Vanderbilt University Medical Center, Nashville, TN, USA*

[3] *Aurum Institute, Johannesburg, South Africa*

*Corresponding author: Qingyan Xiang,
Email: qingyan.xiang@vumc.org; ORCID: 0000-0003-4717-3809

**Abstract**

**Introduction**
Longitudinal lung function recovery after pulmonary disease commonly follows nonlinear trajectories, and failure to adequately model these trajectories can lead to biased or misleading estimates of treatment effects. However, an important methodological gap remains as there is limited assessment of statistical methods for modeling nonlinear lung function trajectories.

**Methods**
We compared several longitudinal modeling approaches for characterizing recovery in percent predicted forced expiratory volume in one second (FEV1p) using data from a phase 2 randomized trial for pulmonary tuberculosis (TB). We estimate the differences in repeated mean FEV1p between each treatment arm and control arm over a 180-days follow-up period. We compared 6 different statistical models: (1) linear mixed-effects model, (2) a piecewise linear mixed-effects model, (3) quadratic and (4) natural cubic spline mixed-effects models, (5) a nonlinear mixed-effects model with exponential recovery function, and (6) a generalized additive mixed model. We discussed the assumptions, clinical interpretations, and resulting treatment-effect estimates across these approaches.

**Results**
The results from the TB trial analyses showed that the conventional linear mixed-effects model provided limited evidence of treatment differences over follow-up, whereas several flexible models identified significant differences during specific periods of recovery.

**Conclusion**
Flexible longitudinal models can complement conventional linear mixed-effects models by revealing treatment differences at certain periods of follow-up that may be obscured by assuming a single linear trend. The choice of nonlinear modeling strategy should be guided by the scientific objective, available data, and the desired balance between clinical interpretability and flexibility.

# 1 Introduction

Chronic and infectious pulmonary diseases, including chronic obstructive pulmonary disease (COPD) and tuberculosis (TB), are major sources of morbidity worldwide and may lead to persistent impairment in respiratory function [1-3]. In clinical studies and trials of pulmonary diseases, lung function measures based on forced expiratory volume in one second (FEV1), including percent predicted FEV1 (FEV1p), are commonly used as outcomes to quantify respiratory impairment and evaluate disease progression or recovery over time [4, 5]. In stable COPD cohorts and long-term trials, FEV1 quantifies gradual chronic decline and is often summarized by an annual slope after any initial treatment response [4, 6]. In contrast, during TB treatment, longitudinal FEV1 reflects recovery from acute lung injury, with the largest gains often occurring early and smaller subsequent changes as healing proceeds [7, 8]. In those studies, longitudinal changes in repeatedly measured FEV1 can provide useful information for characterizing disease trajectories and assessing treatment effects on lung function recovery [8].

Longitudinal lung function outcomes are often summarized using rates of change or analyzed with models that include a linear time trend, implicitly assuming a constant average change in measures such as FEV1 over follow-up [9, 10]. This approach provides a simple and interpretable summary, but it may oversimplify clinical recovery patterns. Lung function recovery after pulmonary injury or infection may be nonlinear, with rapid early improvement followed by plateauing or slower subsequent improvement [7, 11]. In addition, longitudinal lung function patterns may vary substantially across individuals and clinical subgroups [12, 13]. Consequently, models based on a simple linear time trend may obscure clinically meaningful differences between treatment groups, particularly when benefits or harms occur only during specific periods of follow-up. These considerations motivate the use of flexible longitudinal models that can characterize nonlinear recovery trajectories and estimate treatment differences continuously across follow-up.

Several studies have compared or discussed the advantages of using nonlinear methods to model longitudinal outcomes in other clinical domains [14-16]. Meanwhile, flexible models have been used to describe longitudinal lung function patterns in pulmonary diseases [12, 17]. However, the extent to which different longitudinal modeling strategies may produce consistent conclusions about treatment differences in lung function recovery across time, or identify treatment effects that may be missed by simpler approaches, remains unclear. Therefore, in this study, we compared several longitudinal modeling strategies for characterizing nonlinear lung function recovery among participants in a TB host-directed therapy trial, in which adjunctive host-directed therapies were compared with standard TB treatment alone. Specifically, we evaluated linear mixed-effects models, mixed-effects models with polynomial time terms or splines, nonlinear mixed-effects models, and generalized additive mixed models. We used these approaches to estimate longitudinal recovery trajectories in FEV1p and differences between each host-directed therapy arm and standard therapy across follow-up, illustrating how flexible modeling can complement conventional linear approaches in detecting clinically relevant treatment effects over time.

*Table 1*. Baseline characteristics by treatment arm. Continuous variables are summarized as median (interquartile range), and categorical variables are summarized as n (%). Summaries are based on non-missing values.

| | Standard | CC-1105 | Everolimus | Auranofin | Ergocalciferol | Overall |
|---|---|---|---|---|---|---|
| | (N=40) | (N=39) | (N=38) | (N=40) | (N=40) | (N=197) |
| **Baseline FEV1 Percent Predicted** | | | | | | |
| Median (IQR) | 62.0 (20.0) | 63.0 (38.5) | 73.0 (29.0) | 63.0 (29.8) | 69.0 (39.3) | 64.0 (32.0) |
| **Sex** | | | | | | |
| Male | 38 (95.0%) | 32 (82.1%) | 33 (86.8%) | 37 (92.5%) | 33 (82.5%) | 173 (87.8%) |
| Female | 2 (5.0%) | 7 (17.9%) | 5 (13.2%) | 3 (7.5%) | 7 (17.5%) | 24 (12.2%) |
| **Age Group (Years)** | | | | | | |
| 18–24 | 8 (20.0%) | 4 (10.3%) | 6 (15.8%) | 4 (10.0%) | 4 (10.0%) | 26 (13.2%) |
| 25–34 | 15 (37.5%) | 16 (41.0%) | 13 (34.2%) | 11 (27.5%) | 12 (30.0%) | 67 (34.0%) |
| 35–44 | 8 (20.0%) | 14 (35.9%) | 12 (31.6%) | 16 (40.0%) | 12 (30.0%) | 62 (31.5%) |
| 45–64 | 9 (22.5%) | 5 (12.8%) | 7 (18.4%) | 9 (22.5%) | 12 (30.0%) | 42 (21.3%) |
| **BMI** | | | | | | |
| Median (IQR) | 18.5 (3.05) | 18.6 (2.65) | 19.4 (4.00) | 18.4 (2.78) | 18.3 (3.30) | 18.6 (3.23) |
| **Smoking History** | | | | | | |
| No | 15 (37.5%) | 16 (41.0%) | 20 (52.6%) | 20 (50.0%) | 22 (55.0%) | 93 (47.2%) |
| Yes | 25 (62.5%) | 23 (59.0%) | 18 (47.4%) | 20 (50.0%) | 18 (45.0%) | 104 (52.8%) |
| **Radiographic Extent of Disease** | | | | | | |
| Moderately Advanced | 24 (60.0%) | 23 (59.0%) | 26 (68.4%) | 21 (52.5%) | 21 (52.5%) | 115 (58.4%) |
| Far Advanced | 16 (40.0%) | 16 (41.0%) | 12 (31.6%) | 19 (47.5%) | 19 (47.5%) | 82 (41.6%) |
| **Cavity Diameter (cm)** | | | | | | |
| 0 | 11 (27.5%) | 5 (12.8%) | 9 (23.7%) | 2 (5.0%) | 4 (10.0%) | 31 (15.7%) |
| 0–4 | 24 (60.0%) | 23 (59.0%) | 19 (50.0%) | 26 (65.0%) | 28 (70.0%) | 120 (60.9%) |
| ≥4 | 5 (12.5%) | 11 (28.2%) | 10 (26.3%) | 12 (30.0%) | 8 (20.0%) | 46 (23.4%) |
| **Minimum CT Probe Value** | | | | | | |
| Median (IQR) | 16.5 (3.00) | 16.0 (3.50) | 17.0 (5.75) | 17.0 (4.00) | 15.0 (4.25) | 16.0 (5.00) |

# 2 Longitudinal Modeling Methods for Comparison

## 2.1 Illustrative Study

We used the study cohort obtained from a phase 2 randomized controlled trial of adults with pulmonary TB conducted in South Africa as the demonstrating example [8]. A total of 199 participants were randomized in a 1:1:1:1:1 ratio to receive one of four host-directed therapies (HDTs), including CC-1105, Everolimus, Auranofin, and Ergocalciferol, or standard therapy. Pulmonary function tests were performed during the first six months (0–180 days) after randomization. The modified intention-to-treat population, defined as participants with at least one post-baseline pulmonary function test result, was used in all subsequent analyses. Table 1 presents the baseline demographic and clinical characteristics of the 197 eligible participants included in the primary analysis.

The primary outcome was FEV1 measured over follow-up and expressed as a percentage of the predicted value based on age, sex, height, and race, thereby accounting for individual-level differences in lung function (FEV1p) [8]. As illustrated in Figure 1, baseline mean FEV1p differed across arms, with lower values in the Auranofin and standard therapy arms. Over follow-up, mean FEV1p generally increased in all HDT arms except Auranofin. In the standard therapy arm, mean FEV1p initially increased but declined later in follow-up. To characterize how differences between treatment arms evolved over time, the primary estimand was the mean difference in FEV1p between each HDT arm and standard therapy over 0–180 days, which was compared across modeling approaches.

***Fig. 1*** *Treatment-specific mean FEV1p trajectories over follow-up. Points and solid lines show observed mean FEV1p at scheduled visits, and dashed lines show LOESS-smoothed mean trajectories (span = 0.8).*

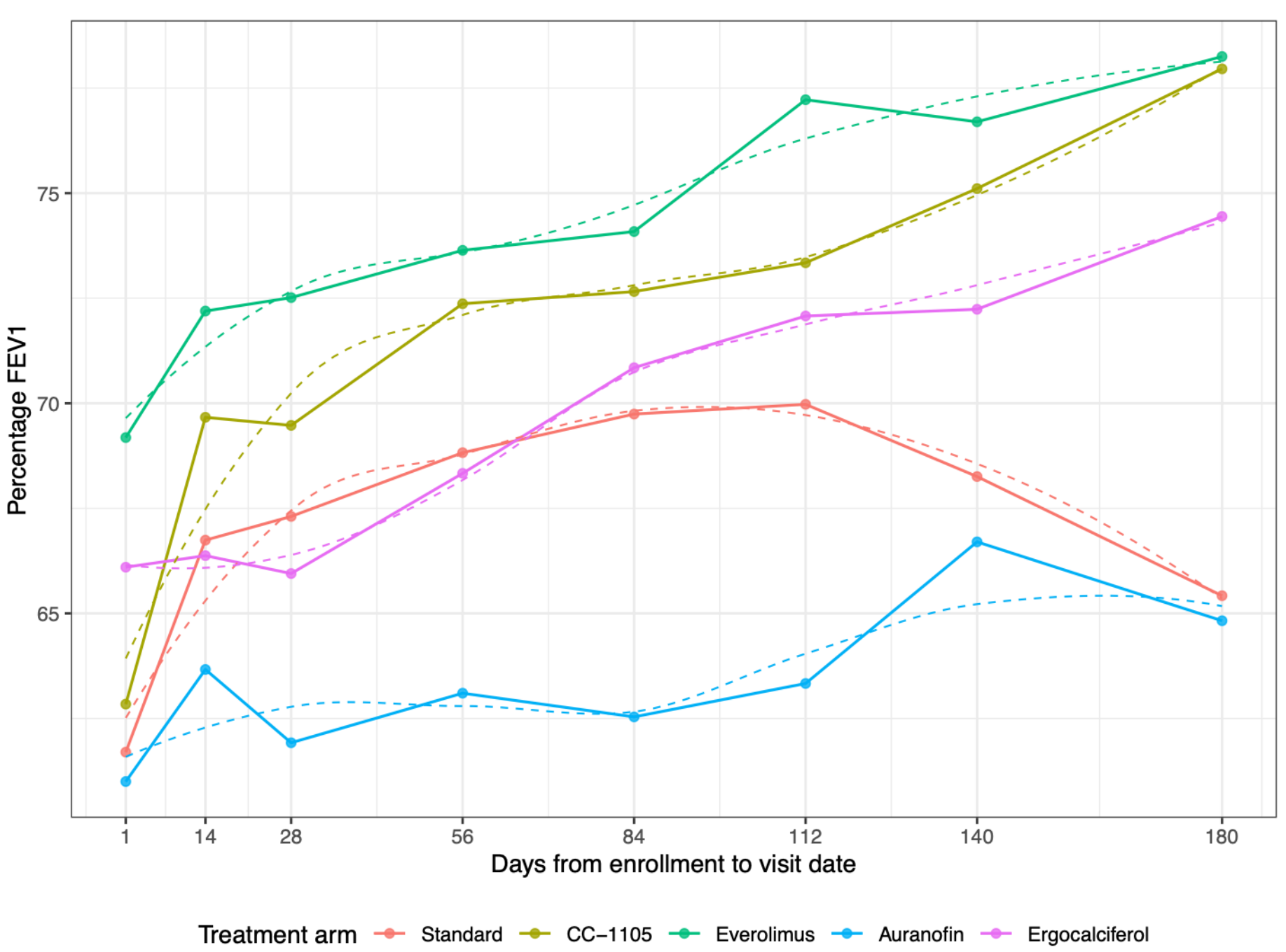

### 2.2 Modeling Methods

We compared several longitudinal modeling approaches to describe FEV1p recovery over follow-up days. All models were fit within a mixed-effects modeling framework, and subject-specific random intercepts were used to account for between-participant differences in baseline FEV1p levels. Each model included treatment arm and treatment-by-time interaction terms or nonlinear extensions to estimate arm-specific recovery trajectories. Throughout the analysis, we adjusted for baseline FEV1p, radiographic severity, minimum CT probe value, and smoking history. To emphasize clinical interpretation, the following sections describe the methods conceptually, and full mathematical specifications are provided in Appendix A.

#### 2.2.1 Baseline: Linear Mixed-Effects Models

We first fit a linear mixed-effects model (LMM) with a linear time trend, which summarizes lung function recovery using a constant average change in FEV1p over the entire follow-up period. This model served as the baseline model for comparison with approaches that allowed more flexible time trends.

#### 2.2.2 Mixed-Effects Models Extensions

Building on the baseline mixed-effects framework, we considered several alternative specifications of follow-up time to capture potential nonlinear recovery patterns. First, we considered a piecewise linear model. In this model, the follow-up period is divided into time intervals by prespecified knots, and the rate of change in FEV1p is allowed to differ across intervals. Therefore, the estimated recovery trajectory is represented by connected linear segments rather than a single straight line over the entire follow-up period. Based on visual inspection of the observed trajectories, we placed one knot at day 30, allowing the average rate of FEV1p change to differ before and after day 30.

Because FEV1p recovery may not follow straight-line patterns even within shorter time intervals, we considered two additional approaches for representing recovery trajectories as smooth curves over time. One approach is to use polynomial models, which allow curvature by adding higher-order terms for time. In this analysis, we extended the baseline linear mixed-effects model to include a quadratic time specification, with both time and time squared. In addition, natural cubic spline models provide a more flexible smooth-curve representation by fitting smooth curves joined at prespecified knots, with additional constraints that improve stability near the boundaries of the observed time range [18]. We fit natural cubic spline models with three degrees of freedom. The quadratic specification and spline degrees of freedom were selected by comparing candidate models using Akaike information criterion (AIC) [19], together with visual assessment of the fitted trajectories.

### 2.2.3 Nonlinear Mixed-Effects Models with Exponential Recovery Function

We also considered a nonlinear mixed-effects model (NLMM) based on a prespecified functional form [20]. Unlike polynomial- or spline-based models, an NLMM allows model parameters to enter the mean function nonlinearly. This approach provides clinically interpretable parameters but imposes stronger assumptions on the shape of the recovery trajectory.

Building on prior evidence that lung function may improve more rapidly in early follow-up and then gradually approach a plateau [7, 11], we considered an exponential recovery function, which properly describes a trajectory aligned with prior evidence. Because the recovery rate enters through the exponential term, the mean function is nonlinear in this parameter. We allowed the total expected improvement to vary across treatment arms to estimate treatment-specific differences in the overall magnitude of recovery, while assuming a common recovery rate.

### 2.2.4 Generalized Additive Mixed Models

NLMMs still require a specific parametric form for the underlying trajectory, which may be restrictive when the true data-generating process deviates from the assumed curve. To further relax this assumption, we also considered generalized additive mixed models (GAMMs). GAMMs extend conventional regression models by replacing selected linear terms with smooth functions estimated from the observed data [21]. In a GAMM, covariate effects, smooth functions of time, and participant-specific random effects are added together in the mean function. The time smooth is constructed as a weighted sum of basis functions, allowing the recovery trajectory to adapt to the data without prescribing a specific parametric shape. In our analysis, treatment-specific smooth functions of follow-up time allowed each treatment arm to have its own nonlinear recovery trajectory.

Table 2 summarizes the modeling approaches and the corresponding assumptions about the shape of FEV1p recovery over time. We fit each modeling approach to the study cohort and compared their results in the following section.

***Table 2**. Summary of longitudinal modeling approaches and their interpretation.*

| Model | Time trend specification | Clinical interpretation |
|---|---|---|
| Linear mixed-effects model | Linear time trend | Assumes FEV1p changes at a constant rate on average over follow-up; the mean trajectory is represented as a straight line. |

| Model | Time trend specification | Clinical interpretation |
| --- | --- | --- |
| Piecewise linear mixed-effects model | Linear spline with one knot at day 30 | Allows the average rate of FEV1p change to differ before and after day 30; the trajectory is represented by two connected straight-line segments. |
| Quadratic mixed-effects model | Linear and quadratic time terms | Allows a smooth curved trajectory, where the rate of FEV1p change can gradually increase or decrease over time. |
| Natural cubic spline mixed-effects model | Spline-based smooth curve | Allows a flexible smooth trajectory over time while improving stability near the beginning and end of follow-up. |
| Nonlinear mixed-effects model | Exponential-like recovery function | Assumes a clinically interpretable recovery pattern with faster early improvement followed by slower improvement or plateauing later. |
| Generalized additive mixed model | Data-driven smooth curve | Allows the most flexible recovery trajectory among the models considered, without prespecifying a particular parametric shape. |

# 3 Results and Model Comparisons

## 3.1 Full Sample Analysis

Figure 2 summarizes the estimated differences in mean FEV1p between each HDT arm and the standard therapy arm over time, together with the corresponding pointwise 95% confidence intervals, across modeling approaches. Under the baseline linear mixed-effects model, the confidence intervals for all four HDT comparisons contained zero throughout follow-up, indicating limited evidence of differences in lung function recovery between any HDT and standard therapy arms.

When more flexible time specifications were incorporated, we identified portions of follow-up during which the pointwise 95% confidence intervals excluded zero. The quadratic and natural cubic spline mixed-effects models suggested higher mean FEV1p for Everolimus compared with standard therapy toward the end of follow-up, and GAMM identified a similar late-follow-up difference. For Auranofin, most nonlinear approaches suggested lower mean FEV1p than standard therapy during mid-follow-up. The piecewise linear model suggested lower mean FEV1p for Ergocalciferol, but this occurred only over a very brief interval near the specified knot and was not reproduced by the other models.

These differences likely reflect how much flexibility each model allows in the shape of lung function trajectory. The piecewise linear model represents the trajectory using connected linear segments, whereas the quadratic, natural cubic spline, and GAMM approaches allow smoother or more flexible curvature. The GAMM, which provides the most flexible and data-adaptive representation among the considered models, additionally suggested a positive difference for CC-1105 toward the end of follow-up. These findings suggest that flexible models may reveal treatment-arm differences that are not captured by simpler specifications.

The nonlinear mixed-effects model with exponential recovery function identified a sustained negative difference for Auranofin after approximately day 60 but did not reproduce the other significant differences observed in other models. This is likely because this model imposed a more restrictive trajectory shape under modestly sized sample, which may not fully capture treatment-specific differences that emerge near the end of follow-up or follow a more complex curved pattern.

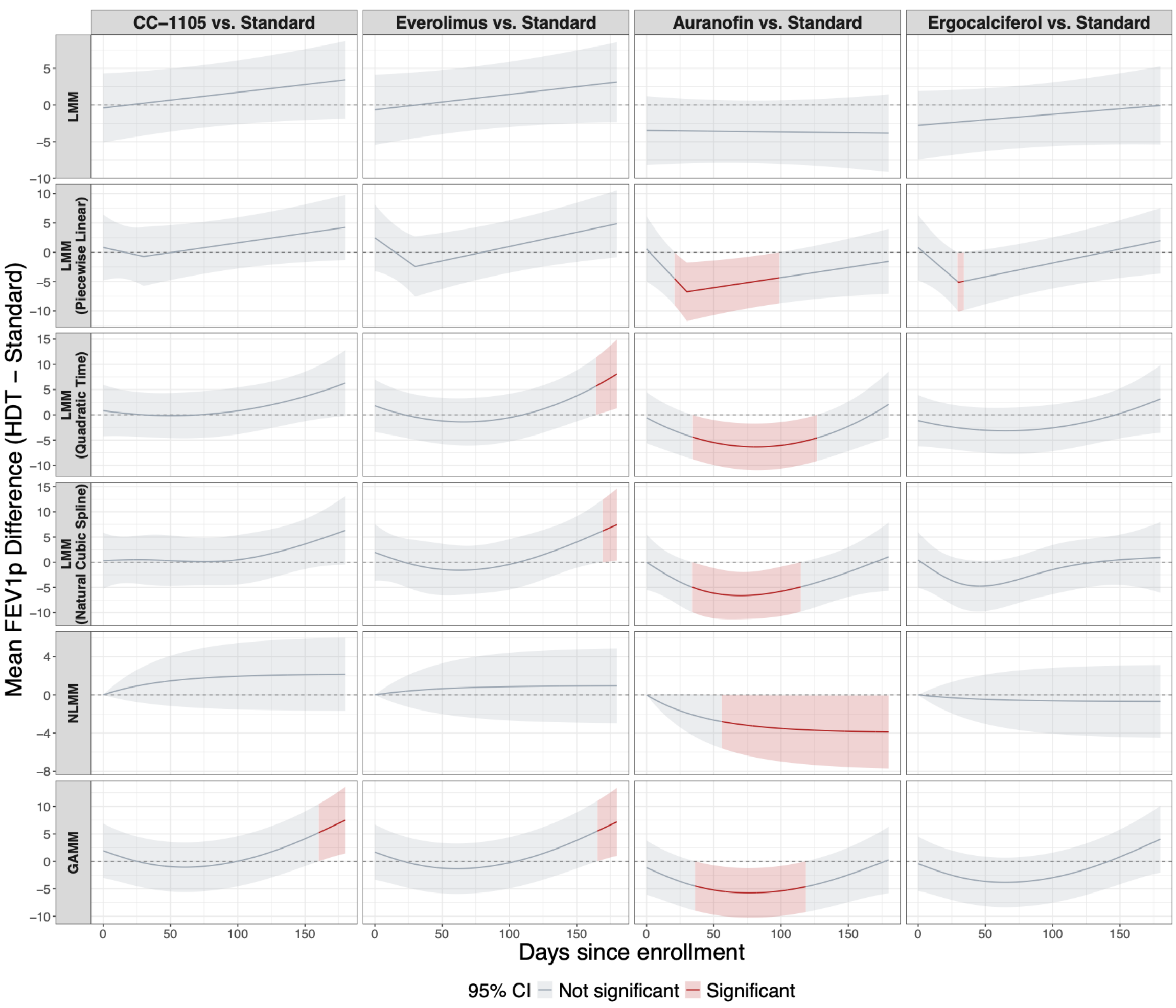

***Fig. 2*** *Differences in mean FEV1p between each HDT and standard therapy over time, with corresponding pointwise 95% confidence intervals, across modeling approaches. Red shaded regions indicate time points at which the pointwise 95 % confidence interval excludes 0.*

### 3.2 Subgroup Analysis of Participants with Moderately Advanced Radiographic Severity

We repeated the same analysis among 115 participants classified as having moderately advanced disease based on baseline chest radiography. The results are summarized in Figure 3. Notably, Everolimus showed a positive difference in mean FEV1p toward the end of follow-up under the baseline linear mixed-effects model and all nonlinear modeling approaches except the nonlinear mixed-effects model. Nevertheless, the nonlinear modeling approaches provided additional insight by identifying a broader portion of follow-up with significant differences.

For Auranofin, the piecewise linear, quadratic, and natural cubic spline approaches identified lower mean FEV1p than standard therapy during mid-follow-up. For CC-1105, the GAMM was the only model that detected a significant positive difference toward the end of follow-up, due to its greater flexibility in capturing changes over time. No clear significant differences were detected for Ergocalciferol in this subgroup.

The nonlinear mixed-effects model with exponential recovery function did not identify significant differences for any HDT arm in this subgroup, further reflecting the limitations of this more restrictive parametric specification in the present data. Overall, among patients with moderately advanced radiographic severity, the later positive difference for Everolimus and the mid-follow-up negative difference for Auranofin were more consistently detected across flexible longitudinal modeling approaches.

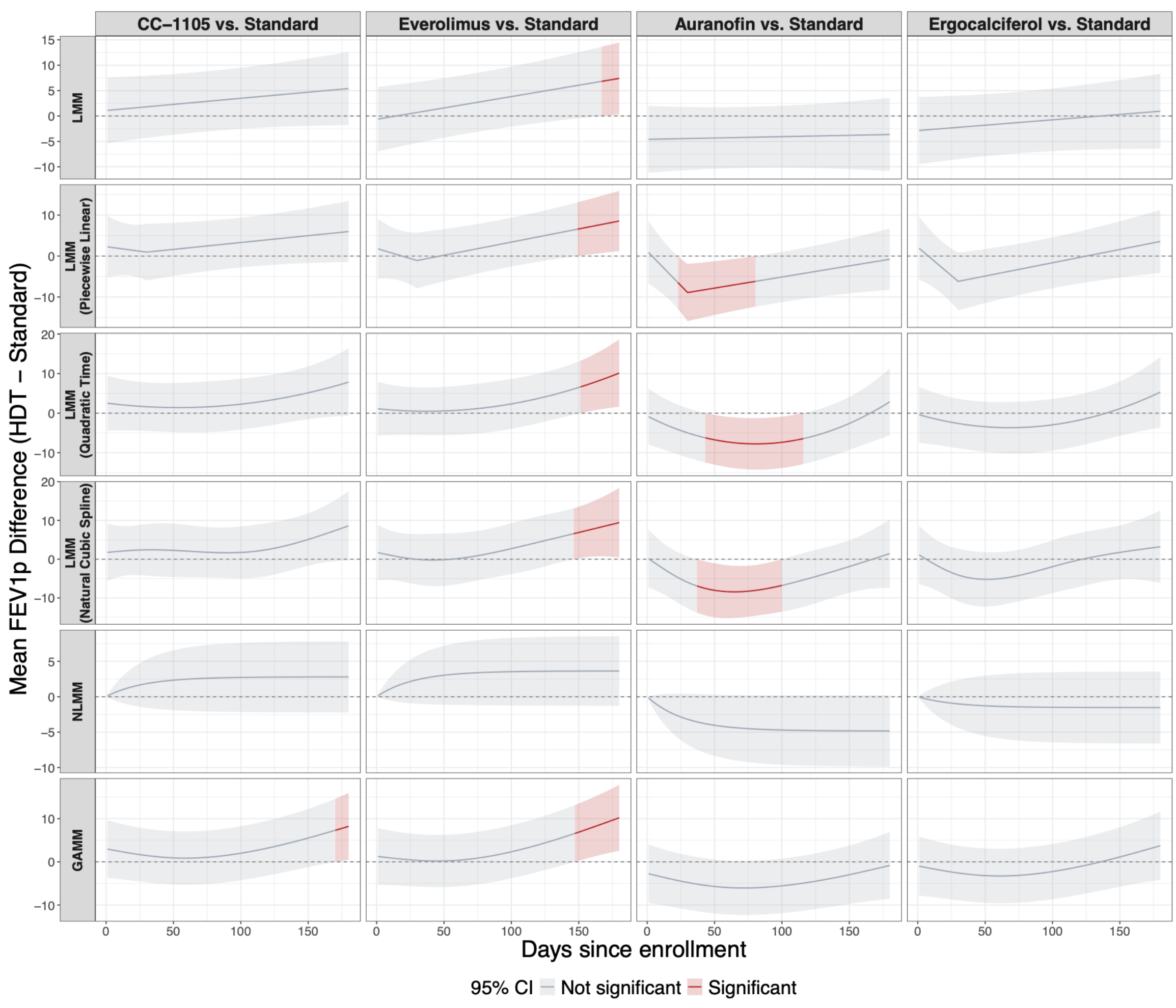

***Fig. 3*** *Differences in mean FEV1p between each HDT and standard therapy over time among patients with moderately advanced radiographic severity, with corresponding pointwise 95% confidence intervals, across modeling approaches. Red shaded regions indicate time points at which the pointwise 95 % confidence interval excludes 0.*

## 4 Discussion

In this study, we compared several longitudinal modeling strategies for characterizing FEV1p recovery after randomization in a phase 2 trial of host-directed therapies for pulmonary tuberculosis. Conclusions about treatment differences in lung function recovery depend on how follow-up time is modeled. Under the baseline linear mixed-effects model, there was limited evidence of differences between any HDT arm and standard therapy across follow-up. Across models that allowed nonlinear recovery, the most consistent patterns were lower mean FEV1p for Auranofin during mid-follow-up and higher mean FEV1p for Everolimus near the end of follow-up. The GAMM alone  suggested a positive difference for CC-1105 toward the end of follow-up in both the full sample and moderately advanced subgroup. Ergocalciferol showed no consistent difference, except for a very brief interval near day 30 under the piecewise linear model. Together, flexible longitudinal models may provide additional insight when treatment effects vary over time and are not well summarized by a single linear trend. However, given the modest sample size, greater model flexibility may increase the risk of overfitting.

These findings have methodological implications for pulmonary trials with repeated lung function outcomes. Linear specifications remain useful because they are simple and interpretable. However, when recovery is nonlinear or treatment effects vary over follow-up, these approaches may obscure certain benefits or harms across time. Flexible longitudinal modeling can complement conventional analyses by estimating treatment differences continuously and more flexibly over time. In early-phase trials, such information may be particularly useful for generating hypotheses, selecting candidate interventions, and informing the timing of outcome assessment in future studies.

Longitudinal lung function has served different purposes in COPD and pulmonary TB studies. COPD trials commonly summarize chronic decline over several years using an annual rate, sometimes estimated after an initial treatment period to distinguish early treatment-related changes from subsequent decline [4, 22, 23]. In contrast, repeated lung function measurements during TB treatments reflect recovery from acute pulmonary impairment, often with relatively rapid early improvement followed by slower or heterogeneous later change [7, 8, 11]. These recovery trajectories remain poorly characterized, as few TB trials have incorporated lung function outcomes [24]. Accordingly, flexible longitudinal models offer an exploratory approach to characterizing lung function recovery and informing future TB trials.

This study has several limitations. First, the study cohort had a relatively small sample size, especially within individual treatment arms and subgroup analyses. The analysis also focused on the first six months after randomization. Larger studies with longer follow-up may provide greater power to detect differences between treatment arms at specific periods, determine whether observed differences persist over time, and support more stable estimation of flexible models.

In addition, this study was intended to illustrate the potential value of nonlinear longitudinal modeling for repeated clinical outcomes, rather than to instruct a single best model. The choice of modeling strategy should be guided by the scientific question, the observed data structure, and clinical knowledge and considerations. When recovery is expected to be approximately linear, simpler models may be sufficient and preferable. When there is clinical or empirical evidence of nonlinear recovery or different treatment effects across time, flexible longitudinal models may provide a useful complementary approach.

**Statements and Declarations**

**Funding:** There was no funding for this work.

**Ethics approval:** This study was a secondary analysis of fully de-identified data. The investigators had no access to participant identifiers; therefore, the analysis did not constitute human subjects research.

**Data availability:** The data are available through the LSHTM Data Compass: https://datacompass.lshtm.ac.uk/id/eprint/2277/

## A Mathematical Representations of Models

### A.1 Baseline Linear Mixed-Effects Model

For the baseline linear mixed-effects model, we specified a linear time trend and treatment-by-time interactions. For participant $i$ at time $j$, the model is specified as

$$\mathrm{FEV1p}_{ij} = \beta_0 + b_i + \beta_1 A_i + \beta_2 T_{ij} + \beta_3 A_i T_{ij} + \boldsymbol{\gamma}^\top \boldsymbol{X}_i + \varepsilon_{ij},$$

where $\mathrm{FEV1p}_{ij}$ denotes the percent predicted FEV1 measurement for participant $i$ at follow-up day $T_{ij}$, and $A_i$ is the dummy variable for treatment indicator. The time-by-treatment interaction, denoted by $A_i T_{ij}$, allows treatment effects to differ in each treatment arm. The matrix $\boldsymbol{X}_i$ included baseline FEV1p, radiographic extent of disease, minimum CT probe value, and smoking history. The subject-specific random intercept $b_i$ accounts for between-participant differences in baseline FEV1p levels and induces correlation among repeated measurements from the same participant. We assumed $b_i \sim N(0, \tau^2)$, the noise variable $\varepsilon_{ij} \sim N(0, \sigma^2)$, and $b_i$ independent of $\varepsilon_{ij}$.

### A.2 Piecewise Linear Mixed-Effects Model

For the piecewise linear mixed-effects model, we have

$$\begin{aligned}\text{FEV1p}_{ij} &= \beta_0 + b_i + \beta_1 A_i + \beta_2 T_{ij} + \beta_3 (T_{ij} - T_0)_+ \\ &+ \beta_4 A_i T_{ij} + \beta_5 A_i (T_{ij} - T_0)_+ + \boldsymbol{\gamma}^\top \boldsymbol{X}_i + \varepsilon_{ij}.\end{aligned}$$

where $(T_{ij} - T_0)_+ = \max(T_{ij} - T_0,\ 0)$, with $T_0 = 30$. This specification allows the slope of FEV1p recovery to change after day 30, with treatment-by-time interaction terms allowing this change to differ by treatment arm.

### A.3 Linear Mixed-Effects Model with Quadratic Time

For the linear mixed-effects model with quadratic time specification, we have

$$\text{FEV1p}_{ij} = \beta_0 + b_i + \beta_1 A_i + \beta_2 T_{ij} + \beta_3 T_{ij}^2 + \beta_4 A_i T_{ij} + \beta_5 A_i T_{ij}^2 + \boldsymbol{\gamma}^\top \boldsymbol{X}_i + \varepsilon_{ij}.$$

Here, the additional quadratic terms allow the average FEV1p trajectory to be represented as a smoothed curve, while the treatment-by-quadratic-time interactions allow the degree of curvature to differ by treatment arm. Clinically, the quadratic time represents the decelerating (or accelerating) improvement of FEV1p over time.

### A.4 Linear Mixed-Effects Model with Natural Cubic Spline

For the linear mixed-effects model with natural cubic spline with three degrees of freedom, we have

$$\text{FEV1p}_{ij} = \beta_0 + b_i + \beta_1 A_i + \sum_{r=1}^{3} \alpha_r\, B_r(T_{ij}) + \sum_{r=1}^{3} \delta_{kr}\, A_i B_r(T_{ij}) + \boldsymbol{\gamma}^\top \boldsymbol{X}_i + \varepsilon_{ij}.$$

Here, $B_r(T_{ij})$ denotes the $r$th natural cubic spline basis function of follow-up time, with three degrees of freedom. The interaction terms between treatment arm and spline basis functions allow each treatment arm to have its own nonlinear recovery trajectory.

### A.5 Nonlinear Mixed-Effects Model with Exponential Recovery Function

For participant $i$ at follow-up time $T_{ij}$, the nonlinear mixed-effects model was specified as

$$\text{FEV1p}_{ij} = y_{0i} + b_i + D_{A_i}\left(1 - e^{-kT_{ij}}\right) + \varepsilon_{ij},$$

where $A_i$ denotes the treatment arm for participant $i$. Here, $y_{0i}$ represents the participant-specific baseline mean FEV1p as a function of baseline covariates, $b_i \sim N(0, \tau^2)$ accounts for between-participant heterogeneity in baseline lung function, and $\varepsilon_{ij} \sim N(0, \sigma^2)$ denotes the residual error. The parameter $D_{A_i}$ represents the magnitude of long-term improvement in FEV1p for treatment arm $A_i$, corresponding to the difference between baseline and the asymptotic plateau level. The recovery rate $k > 0$ determines how quickly lung function approaches the plateau over time. The recovery rate was assumed to be common across treatment arms, because estimating multiple arm-specific nonlinear parameters could lead to convergence difficulties or unstable estimates in this modestly sized sample. Together, $D_{A_i}$ and $k$ determine the shape of the recovery curve, with larger values of $D_{A_i}$ corresponding to greater long-term improvement and larger values of $k$ corresponding to faster early recovery.

### A.6 Generalized Additive Mixed Model

For the generalized additive mixed model, we have

$$\text{FEV1p}_{ij} = \beta_0 + b_i + \beta_1 A_i + f_A(T_{ij})A_i + \boldsymbol{\gamma}^\top \boldsymbol{X}_i + \varepsilon_{ij}.$$

where $f_A(\cdot)$ denotes the arm-specific smooth function of follow-up time. This specification allows treatment-specific recovery trajectories to be estimated flexibly from the observed data.